\documentclass[aps,pra,twocolumn,longbibliography,groupedaddress,showpacs,floatfix,superscriptaddress]{revtex4-1}
\usepackage{mathtools,amssymb,graphicx,units}
\usepackage[plainpages=false,pdfpagelabels,colorlinks=true,linkcolor=red,urlcolor=blue,citecolor=blue,pdftitle={Title},pdfauthor={},pdfdisplaydoctitle=true,pdfduplex=DuplexFlipLongEdge]{hyperref}
\usepackage{soul} 
\usepackage[dvipsnames,usenames]{color}
\usepackage[normalem]{ulem}
\usepackage{color}

\emergencystretch=\maxdimen
\begin{document}

\normalem

\title{Dynamical resilience to disorder: the dilute Hubbard model on the Lieb lattice}
\author{L. Oliveira-Lima} 
\affiliation{Departamento de F\'isica, Universidade Federal do Piau\'i, 64049-550
Teresina PI, Brazil}
\author{N. C. Costa}
\affiliation{Instituto de F\'isica, Universidade Federal do Rio de
Janeiro Cx.P. 68.528, 21941-972 Rio de Janeiro RJ, Brazil}
\affiliation{International School for Advanced Studies (SISSA),
Via Bonomea 265, 34136, Trieste, Italy}
\author{J. \surname{Pimentel de Lima}} 
\affiliation{Departamento de F\'isica, Universidade Federal do Piau\'i,
64049-550
Teresina PI, Brazil}
\author{R. T. Scalettar}
\affiliation{Department of Physics, University of California, Davis, CA 95616,
USA}
\author{R. R. dos Santos} 
\affiliation{Instituto de F\'isica, Universidade Federal do Rio de
Janeiro Cx.P. 68.528, 21941-972 Rio de Janeiro RJ, Brazil}
\begin{abstract}
In itinerant systems, electron-electron interactions may lead to the formation of local magnetic moments and their effective exchange coupling, which in turn gives rise to  long-range magnetic order. 
Therefore, when moment formation is weakened, such as in the single-band Hubbard model on a square lattice with the on-site repulsion being randomly switched off on a fraction $x$ of sites, magnetic order is suppressed beyond some critical $x_c$, which was found to lie below the classical percolation threshold, $x_c^\text{(perc,sq)}$.
Here
we study 
dilute magnetism in flat band systems, namely in the Hubbard model on a `Lieb' lattice.
Interestingly, we show that magnetic order persists to $x$ almost twice as large as the classical percolation threshold for the lattice, thus emphasizing the central role of electron itinerancy to the magnetic response.
The analysis of the orbital-resolved order parameters reveals that the contribution of the four-fold coordinated `d' sites to magnetism is dramatically affected by dilution, while the localized `p' states of the flat band provide the dominant contribution to long-range correlations. 
We also examine the transport properties, which suggest 
the existence of an insulator-to-metal transition in the same range of the critical magnetic dilution.
\end{abstract}

\date{\today}

\maketitle

\section{Introduction}

The study of magnetic systems with quenched random site or bond dilution has raised fundamental issues over the years.
One question which was the subject of considerable scrutiny was whether or not critical exponents are altered;
the Harris criterion suggests they remain at pure system values as long as
the specific heat exponent $\alpha>0$~\cite{harris74}.
Initial explorations focused on classical spin models. 
For instance,
Monte Carlo simulations of
the square lattice bond-diluted~\cite{zobin78} and
site-diluted~\cite{martins07} Ising models
verified `strong universality'.
The exponents were found to be
the same as that of the pure system (even though $\alpha=0$).
Another issue of special experimental interest relates to dilute magnetic semiconductors,
since even a few of
percent transition-metal atoms introduce ferromagnetism, which might be
harnessed to change device functionality~\cite{wolf01,zutic04}.  The
possibility of magnetic order even in the high dilution limit emphasizes
the crucial role of coupling of moments through the free carriers, an
effect not captured in Ising-like models of spins interacting purely
through local exchange coupling.

Indeed,
in dilute magnetic {\it insulators} where there is no long-range 
Rudermann-Kittel-Kasuya-Yosida (RKKY)~\cite{ruderman54,kasuya56,yosida57} 
coupling and interactions
are only between neighboring sites,
geometrical aspects dominate the
suppression of magnetic order as the
interactions between the localized spins are randomly switched off.
This can be achieved by either replacing atoms possessing localized moments by
non-magnetic atoms (the site-dilution problem), or by removing atoms
mediating the superexchange interaction between localized spins
(the bond-dilution problem)~\cite{Stinchcombe83,Belanger00}.  In both
cases, the underlying lattice structure is fundamental,
since only in the
percolating regime in which at least one path of
connected sites spans the whole lattice~\cite{Stauffer94} can long-range
magnetic order be established.  The ground state magnetization
decreases steadily and vanishes at some critical concentration of sites
(s) or bonds (b), such that $x_c^\text{(s)}\leq x_c^\text{(b)}$
\cite{Stauffer94}, beyond which no long-range order can be sustained
\cite{Stinchcombe83,Belanger00}.  


For itinerant systems, however, the situation is quite different.
Consider the repulsive Hubbard model in which the on-site interaction
$U$ is switched off on a fraction $x$ of sites.
Ulmke et
al.\,\cite{Ulmke98b} considered a square lattice at half-filling and
ratio of on-site interaction to hopping integral $U/t=8$.
Long range
antiferromagnetic (AF) order disappears at $x_c \gtrsim 0.43 \pm 0.07$.
The large uncertainty results from the challenges in doing the finite
size and zero temperature extrapolations.  Nevertheless, this strong
coupling critical value is
consistent with the classical site-percolation threshold,
$x_c^\text{(perc,sq)}=0.41$~\cite{Stauffer94}.  
On the other hand, at weaker coupling $U/t=4$,
deviations from classical percolation have been found
\cite{Litak00,Hurt05,Pradhan18,Mondaini08}.
For the square lattice
$x_c$ is significantly less than the expected percolation value
\cite{Hurt05,Mondaini08}
\footnote{These Quantum Monte Carlo studies were actually done on
superconductivity and charge order in
the attractive Hubbard model, but the conclusions for magnetism in
the repulsive case can be obtained through a particle-hole
transformation~\cite{Robaszkiewicz81,dosSantos93}.}.

The fact that,  at coupling $U/t=4$,
the dilution threshold for itinerant electrons
is lower than the classical, geometry-dependent, percolation value suggests
that enhanced double occupancy plays a role in weakening magnetic
order before the percolation threshold is reached.
Interestingly, the
recovery of the percolation value at $U/t \sim 8$
is consistent with the fact that
this is the crossover interaction strength to the regime where the Hubbard model
is well described by the Heisenberg Hamiltonian 
\cite{staudt00,kozik13,khatami16}.
That is, charge fluctuations are strongly suppressed
and no longer play a role
on the effects of dilution in the eventual magnetic response.


In this paper we extend this understanding of dilution in itinerant
electron system to the Hubbard model on the Lieb lattice
\cite{Noda09,Weeks10,Zhao12,Nita13,Costa16,Bercx17,Kumar17,Jeong19,Le19}, also known as a decorated square
lattice, or as the `CuO$_2$ lattice'; see Fig.\,\ref{fig:Lieb}\,(a).
This geometry is 
realized in the CuO$_2$ sheets of high-$T_c$ cuprates 
\footnote{
In this situation $U_d$ on the copper orbitals
is quite a bit larger than $U_p$ on the oxygen orbitals.
}
and also has been emulated in photonic and optical lattices
\cite{Shen10,GuzmanSilva14,Mukherjee15,Vicencio15,Taie15,Xia16,Diebel16},
as well as in atomic manipulation of electronic states in Cu(111) surfaces \cite{Slot17}.  
The Lieb lattice geometry 
allows us to explore diluted itinerant electron systems in
an entirely new physical context, one in which a 
flat band is present at half-filling (for
the noninteracting case),
as displayed in Fig.\,\ref{fig:Lieb}\,(c),
and for which compact localized states 
are present even for
strong hopping disorder \cite{Ramachandran17}.  
As a consequence of these features, the electron
dynamics on the Lieb lattice is quite different from that on 
more conventional structures, 
leading to the possibility of alternate magnetic response when
electron-electron interactions are taken into account.  Indeed, 
the fact that the two sublattices have unequal numbers of sites 
already gives rise 
to unique physics even in  the absence of dilution: a 
\emph{ferrimagnetic} state at half-filling 
\cite{Lieb89,Lieb89err}, with a large
contribution of the $p$-band to this long-range ordered state
\cite{Costa16}.  Here we investigate the robustness of this
ferrimagnetic state in presence of site disorder.  
However, from the outset we stress that due to limitations on the system
sizes used for quantum systems~\cite{Sandvik02a,Sknepnek04,Peng19}, in particular to the itinerant electronic case,
numerical calculations can rarely
provide critical exponents with sufficient accuracy to settle
issues related to the Harris criterion. So, although we have discussed
this issue to lend broad perspective to our work, we will not attempt
to address this issue directly.

The layout of the paper is as follows.  
In Sec.\,\ref{sec:HQMC}
we present details of the model, the calculational procedure,
determinant Quantum Monte Carlo (DQMC),
and the magnetic and transport observables used to
characterize the system.  
The
results are presented and discussed in Sec.\,\ref{sec:results}, while
Sec.\,\ref{sec:conc} summarizes our findings.

\section{Model and Methodology}
\label{sec:HQMC}

The Hubbard Hamiltonian for the Lieb lattice reads
\begin{align} \label{eq:hamiltonian}
\widehat{\mathcal{H}} = & -t_{pd}\sum_{\mathbf{r}\sigma}\Bigl(d^{\dagger}_{\mathbf{r}\sigma}p^{x}_{\mathbf{r}\sigma} + d^{\dagger}_{\mathbf{r}\sigma}p^{y}_{\mathbf{r}\sigma} + \mathrm{H.c}\Bigr) \nonumber \\
& -t_{pd}\sum_{\mathbf{r}\sigma}\left(d^{\dagger}_{\mathbf{r}\sigma}p^{x}_{\mathbf{r} + \widehat{x}\sigma} + d^{\dagger}_{\mathbf{r}\sigma}p^{y}_{\mathbf{r} + \widehat{y}\sigma} + \mathrm{H.c}\right) \nonumber \\
& + \sum_{\mathbf{r},\alpha}U^{\alpha}_{\mathbf{r}}\left(n^{\alpha}_{\mathbf{r}\uparrow} - \frac{1}{2}\right)\left(n^{\alpha}_{\mathbf{r}\downarrow} - \frac{1}{2}\right) \nonumber \\
& + \sum_{\mathbf{r},\sigma,\alpha}(\varepsilon_{\alpha}-\mu)n^{\alpha}_{\mathbf{r}\sigma}  \,\, ,
\end{align}
with $d_{\mathbf{r}\sigma}$, $p^{x}_{\mathbf{r}\sigma}$, and
$p^{y}_{\mathbf{r}\sigma}$ being standard annihilation electron
operators in second-quantized formalism, while
$n^{\alpha}_{\mathbf{r}\sigma}$ are the number operators for their
corresponding orbitals, $\alpha =d$, $p^{x}$, or $p^{y}$; our notation
therefore follows closely that of the CuO$_2$ lattice realization.
The first two terms on the right hand side
of Eq.\,\eqref{eq:hamiltonian} denote the inter- and intra-cell hopping
between $d$- and $p$-orbitals, respectively, while the third term
corresponds to a site and orbital-dependent local repulsive interaction.  The last term
involves the onsite energies $\varepsilon_{\alpha}$ and the chemical
potential $\mu$, which we set to $\varepsilon_{\alpha}=\mu=0$,
a choice which makes each orbital precisely
half-filled.  The hopping integral is taken as $t_{pd}=1$, thus
defining the energy scale.

\begin{figure}[t]
\centering
\includegraphics[scale=0.3]{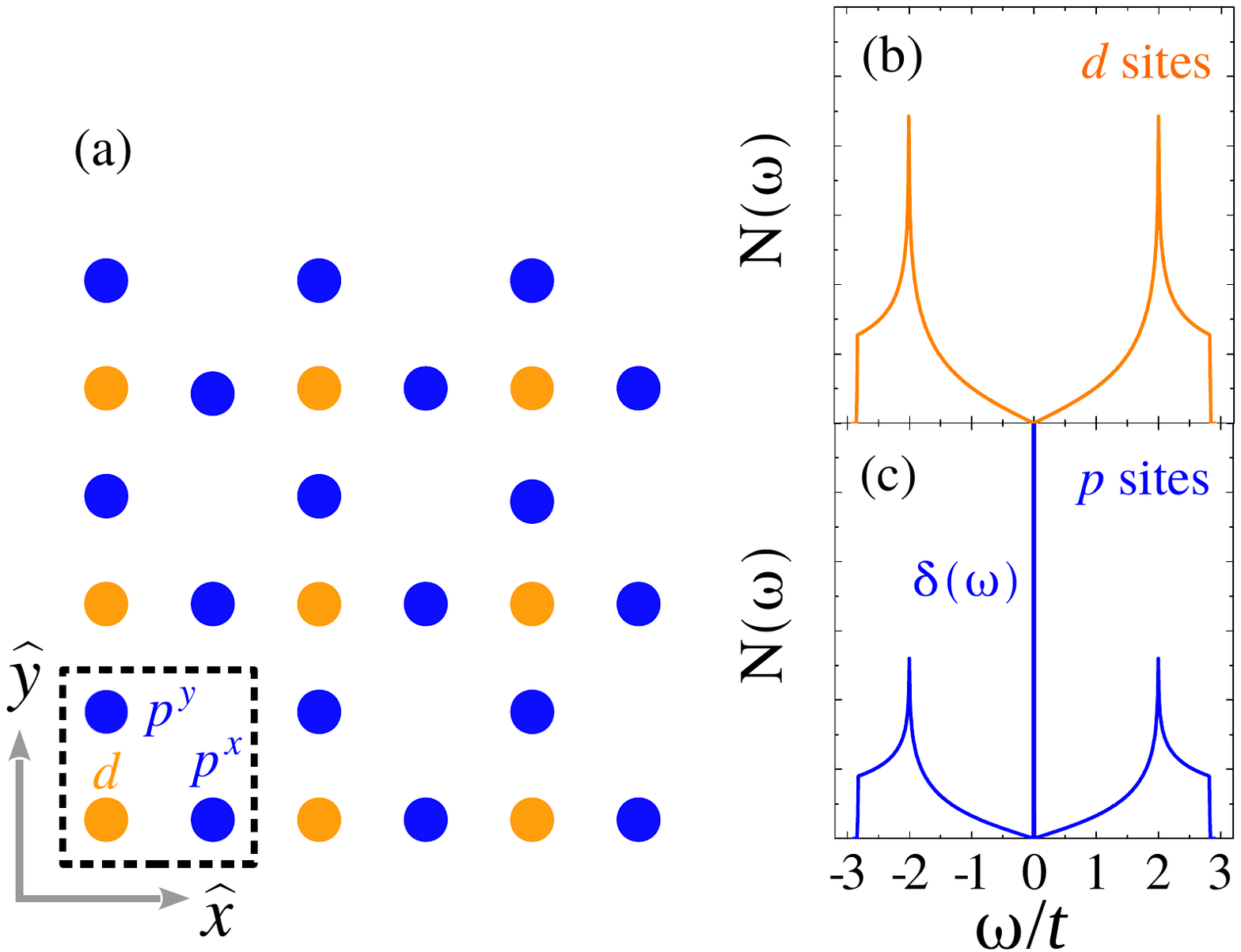}
\caption{(Color online) (a) The Lieb (or CuO$_2$) lattice. The fourfold
coordinated $d$ sites appear in lighter color (orange) and belong to one
sublattice, while the twofold coordinated $p$ sites appear in darker
color (blue) and belong to the other sublattice.
The dashed box corresponds to the unit cell.
Panels (b) and (c) respectively show the non-interacting density of states on $d$ and $p$ sites.
}
\label{fig:Lieb} 
\end{figure}

We model dilution by allowing for random distributions of
$U^{\alpha}_{\mathbf{r}}$, such that a fraction $x$ of
the sites have their interaction strength suppressed,
\begin{equation}
U^{\alpha}_{\mathbf{r}}= \left \{
    \begin{array}{c l}	
         U & {\rm with\,\,probability}\,\,(1-x); \\
         0 & {\rm with\,\,probability}\,\,x .
    \end{array}\right.
\end{equation}
The $U=0$ sites no longer support
moment formation as a result of charge fluctuations.  
Our simulations focus on the intermediate coupling value 
$U/t_{pd}=4$, since this is the case where previous work has found
that magnetic order vanishes (below)
away from the percolation value.
It is important to notice that, for the noninteracting case, since the bandwidth of the Lieb lattice is $W_{\rm lieb}/t =4\sqrt{2} $ [see, e.g., Fig.\,\ref{fig:Lieb}\,(b)-(c)], being smaller than the one for the square lattice ($W_{\rm sqr}/t =8 $), we effectively have a larger $U/W$ for the former.
Therefore, one should naively expect stronger geometrical effects for this choice of interaction strength.

We investigate the ground state properties of the Hamiltonian
\eqref{eq:hamiltonian} by means of DQMC simulations
\cite{Blankenbecler81,Hirsch83,Hirsch85,White89,dosSantos03b}.  This is
an unbiased numerical method based on an auxiliary-field decomposition
of the interaction which maps onto free fermions moving
in a fluctuating space and (imaginary) time dependent potential.
The first key step is a separation (the Trotter-Suzuki decoupling) of the
noncommuting parts of the Hamiltonian: $\widehat{\mathcal{H}}_{0}$
containing the terms quadratic in the fermion creation and 
destruction operators and the quartic term 
$\widehat{\mathcal{H}}_{\rm U}$ which occur in the partition function
$\mathcal{Z}= \mathrm{Tr}\,
e^{-\beta\widehat{\mathcal{H}}}= \mathrm{Tr}\,
[(e^{-\Delta\tau(\widehat{\mathcal{H}}_{0} + \widehat{\mathcal{H}}_{\rm
U})})^{M}]\thickapprox \mathrm{Tr}\,
[e^{-\Delta\tau\widehat{\mathcal{H}}_{0}}e^{-\Delta\tau\widehat{\mathcal{H}}_{\rm
U}}e^{-\Delta\tau\widehat{\mathcal{H}}_{0}}e^{-\Delta\tau\widehat{\mathcal{H}}_{\rm
U}}\cdots]$, where $\beta=M \Delta\tau$, with $\Delta\tau$ being the
grid of the imaginary-time coordinate axis.  This decomposition leads to
an error proportional to $(\Delta\tau)^{2}$, which can be 
systematically reduced as $\Delta\tau
\to 0$.
Here, we choose $\Delta\tau=0.125$, which is small enough so
that the systematic errors
for the magnetic structure factor
are comparable to the statistical ones (from the
Monte Carlo sampling).  The second central step is
a discrete Hubbard-Stratonovich
(HS) transform \cite{Hirsch83} on the two-particle terms
$e^{-\Delta\tau\widehat{\mathcal{H}}_{\rm U}}$ which converts them
also to quadratic in the fermion operators. 
In this way the resulting trace of
fermions propagating in an auxiliary
bosonic field, whose components depend on the space and imaginary-time
lattice coordinates, can be performed.

The HS fields are sampled by
standard Monte Carlo techniques, allowing the measurement of Green's
functions, and other physical quantities including spin, charge, and
pair correlation functions.  The DQMC method,
as with many fermionic QMC approaches, 
in general suffers from the minus-sign
problem when particle-hole symmetry (PHS) is broken
\cite{Troyer05}.  
Here, however, we stress that the Lieb lattice is bipartite and the
introduction of randomness in the interaction strength preserves PHS at
half-filling, so that the sign problem is absent
for this case. 
A detailed introduction to DQMC can be found, e.g.\ in
Refs.\,\onlinecite{dosSantos03b,assaad02,gubernatis16}.  

\begin{figure}[t]
\centering
\includegraphics[scale=0.40]{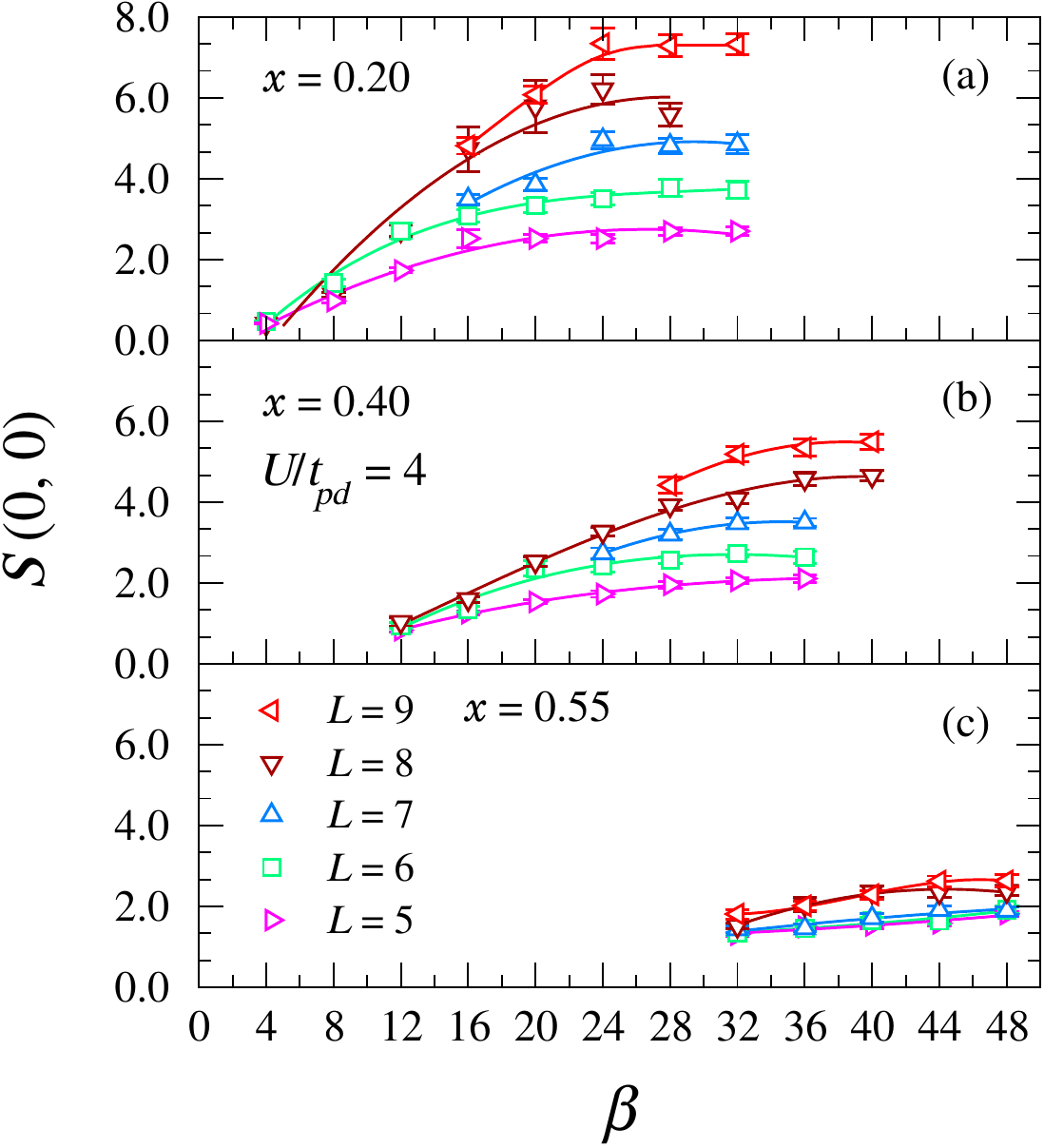}
\caption{(Color online) The global FM spin structure factor as a function
of inverse of temperature for (a) $x=0.20$, (b) $x=0.40$, (c) $x=0.55$,
and different lattice sizes $L$.  Solid lines are guides to the eye.
Here, and in all subsequent figures, when not shown, error bars are
smaller than the symbol size.
}
\label{fig:S00beta} 
\end{figure}

The magnetic response of the system is probed by the real space spin-spin correlation functions
\begin{equation}\label{Eq:Spincorrelation}
c_{\alpha\gamma} (\boldsymbol\ell)= \frac{1}{3} \langle \mathbf{S}^{\alpha}_{\mathbf{r}_{0}} \cdot \mathbf{S}^{\gamma}_{\mathbf{r}_{0} + \boldsymbol\ell} \rangle \, ,
\end{equation}
with $\mathbf{r}_{0}$ being the position of a given unit cell, while $\alpha$ and $\gamma$ denote the orbitals ($d$, $p^{x}$, or $p^{y}$).
The Fourier transform of $c_{\alpha\gamma}(\boldsymbol\ell)$ is the magnetic structure factor,
\begin{equation}\label{Eq:StrucFactor}
S(\mathbf{q}) = \frac{1}{N_{s}} \sum_{\alpha, \gamma}\sum_{\boldsymbol\ell}c_{\alpha\gamma}(\boldsymbol\ell)e^{{\rm i}\mathbf{q}\cdot\boldsymbol\ell} \, ,
\end{equation}
where the number of sites is $N_s=3L^2$, with $L$ being the linear size of the underlying Bravais square lattice. 
$S(\mathbf{q})$ peaks at the dominant magnetic wavevector of the system.
The existence of a global 
\emph{ferro}magnetically ordered state is probed by the usual Huse finite-size scaling form \cite{Huse88}  with $\mathbf{q}=(0,0)$,
\begin{equation}
	\frac{S(0,0)}{L^2}=(m^F)^2+\frac{A}{L},
\label{eq:Huseglobal}	
\end{equation}	 
where $m^F$ is the associated order parameter, and $A$ is a constant.

In addition, for a global ferromagnetic arrangement,
Eq.\,\eqref{Eq:StrucFactor} allows us to separate the individual orbital
contributions as
\begin{align}
\nonumber S(0,0) & = \big( S_{d\,d} + S_{p^{x}\,p^{x}} + S_{p^{y}\,p^{y}} \\
& + 2S_{p^{x}\,p^{y}} + 2S_{d\,p^{x}} + 2S_{d\,p^{y}} \big) / 3 \, ,
\end{align}
with
\begin{align}
S_{\alpha\gamma} = \frac{1}{L^2} \sum_{\boldsymbol\ell}c_{\alpha\gamma}(\boldsymbol\ell) \, ,
\end{align}
where we use the fact that $S_{\alpha\gamma}=S_{\gamma\alpha}$.  Since
the $\pi/2$ real space rotational invariance is recovered after disorder
averaging, one should find 
$S_{p^{x}\,p^{x}} = S_{p^{y}\,p^{y}} \approx S_{p^{x}\,p^{y}}$, and
$S_{d\,p^{x}} = S_{d\,p^{y}}$.  It is therefore useful to define
\begin{align}
S_{p\,p} = \frac{1}{4} \big[ S_{p^{x}\,p^{x}} + S_{p^{y}\,p^{y}} + 2S_{p^{x}\,p^{y}} \big] \, ,
\end{align}
and 
\begin{align}
S_{d\,p} = \frac{1}{2} \big[ S_{d\,p^{x}} + S_{d\,p^{y}} \big] \, ,
\end{align}
which leads to
\begin{align}\label{Eq:StrucFactor2}
S(0,0) & = \frac{1}{3} \big[ S_{d\,d} + 4 S_{p\,p} - 4 |S_{d\,p}| \, \big] \, .
\end{align}
The last term in Eq.\,\eqref{Eq:StrucFactor2} enters in absolute value
because the $d$-$p$ spin correlations are always antiferromagnetic at
half-filling, i.e.\,$S_{d\,p}<0$ (in accordance to rigorous results
derived in Ref.\,\onlinecite{shen94}), and indicative of the
\emph{ferri}magnetic nature which combines a nonzero overall ferromagnetism 
with anti-alignment of $d$ and $p$ spins within the unit cell.

The individual components of the structure factors obey the
same finite size scaling form\cite{Huse88}, allowing us to extract the
orbital-resolved order parameters in the thermodynamic limit,  
\begin{align}\label{Eq:individuals2}
(m^{F}_{d\,d})^{2} & = \frac{S_{d\,d}}{L^2} + \frac{a}{L} \, , \\
 (m^{F}_{p\,p})^{2} & = \frac{S_{p\,p}}{L^2} + \frac{b}{L} \, , \\
(m^{AF}_{d\,p})^{2} &= \frac{|S_{d\,p}|}{L^2} + \frac{c}{L} \,,
\end{align}
where $a$, $b$ and $c$ are constants.

Finally, the metallic or insulating character of the system is probed with two independent quantities. 
One is the direct-current conductivity,
\begin{equation}\label{eq:sigma_dc}
\sigma_{dc} = \frac{\beta^2}{\pi} \Lambda_{xx}(\mathbf{q=0}, \tau = \beta/2),
\end{equation}
where
\begin{equation}
\Lambda_{xx}(\mathbf{q}, \tau ) = 
\langle j_{x}(\mathbf{q}, \tau) j_{x}(-\mathbf{q}, 0)  \rangle,
\end{equation}
with $ j_{x}(\mathbf{q}, \tau) $ being the Fourier transform of
\begin{equation}
j_x(\mathbf{i},\tau)=\mathrm{e}^{\tau\mathcal{H}}
  \left[
        it\sum_\sigma
            \left(c_{\mathbf{i}+\mathbf{x}\sigma}^\dagger 
                  c_{\mathbf{i}\sigma}^{\phantom{\dagger}}
                  - 
                  c_{\mathbf{i}\sigma}^\dagger  
                  c_{\mathbf{i}+\mathbf{x}\sigma}^{\phantom{\dagger}}
            \right)
  \right]
\mathrm{e}^{-\tau\mathcal{H}};
\label{jx}
\end{equation}
This approximation has been extensively used to identify metal-insulator transitions
\cite{Trivedi96,Mondaini12}.
The other quantity is the electronic compressibility, defined as
\begin{equation}
	\kappa = \frac{1}{n^2}\frac{\partial n}{\partial \mu},
\label{eq:kappa}
\end{equation}
where $n$ is the global electronic density.
We should note that in principle our data for $\kappa$ could
suffer from the sign-problem when the chemical potential moves slightly
around half-filling in the finite difference implementation of
Eq.\,\eqref{eq:kappa}.
However, in a regime where $\kappa$ is
small, the sign problem is less serious, because even though
a non-zero chemical potential is applied, the density stays close to
half-filling.
Indeed, we have systematically checked that the average sign
was always close to 1 within the range of parameters analyzed.
Therefore, our data for the compressibility are free from the minus-sign
problem.

\section{Results}
\label{sec:results}

\begin{figure}
\centering
\includegraphics[scale=0.28]{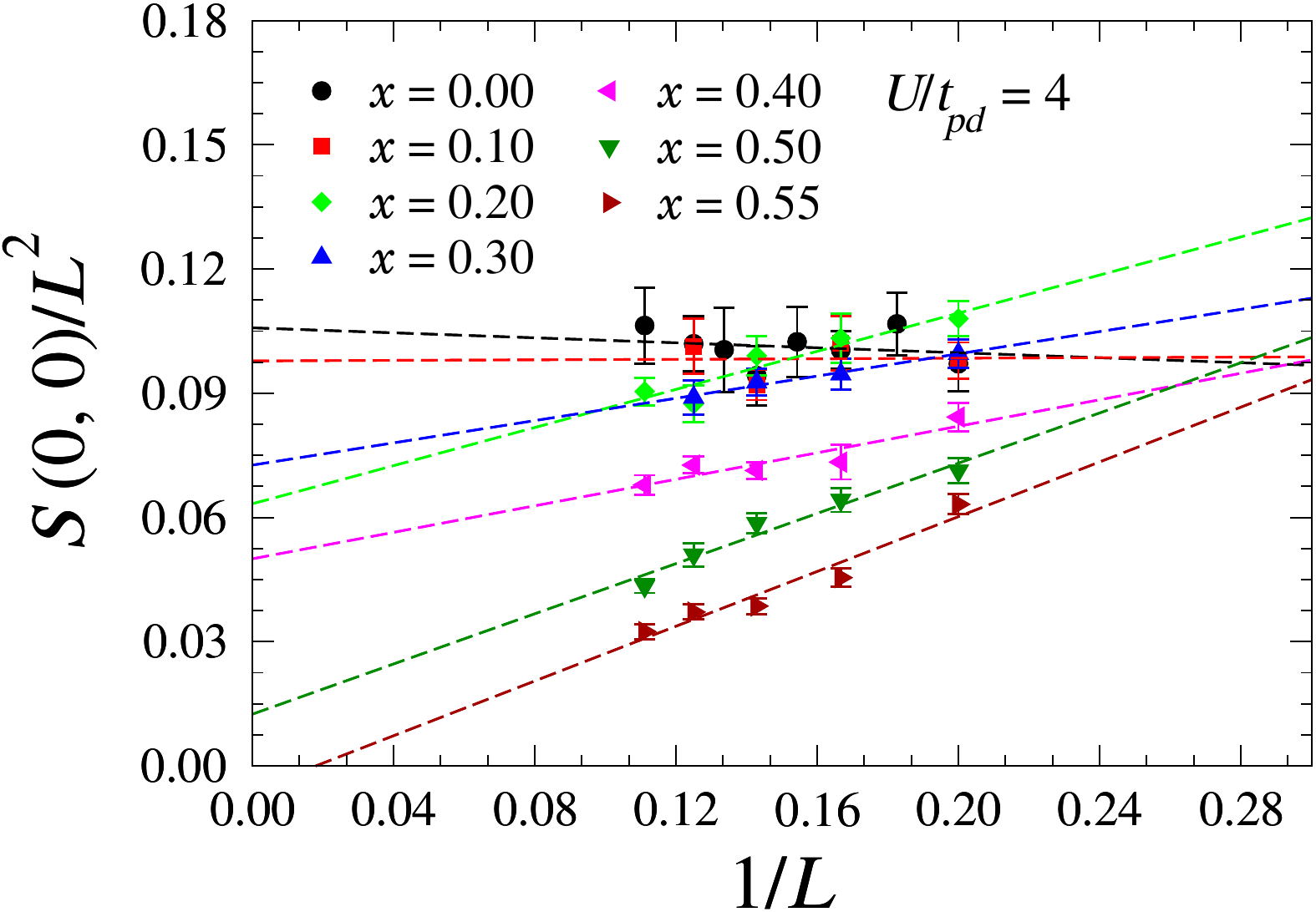}
\caption{(Color online) Finite-size scaling of the normalized 
global ground state structure factor for different impurity concentrations. 
}
\label{fig:S00Huse} 
\end{figure}

We consider lattices with linear sizes up to $L=9$ ($N_{s}\leq 243$),
and we take $U/t_{pd}=4$ throughout the paper.
Further, in what follows our results are obtained by averaging over 20-60 disorder realizations, depending on the temperature and lattice sizes;
this procedure keeps the error bars in the correlation functions small enough to give rise to unambiguous extrapolations.
The disorder configurations are generated in a canonical ensemble,
i.e.~for a given concentration, $x$, of free sites one randomly chooses
$xN_s$ sites to set $U=0$, so that there are no fluctuations in the
number of free sites.  
For dilution fractions $x$
which do not correspond to an integer number of sites for a given $L$,
we perform a weighted average over the adjacent integers.

\begin{figure}
\centering
\includegraphics[scale=0.28]{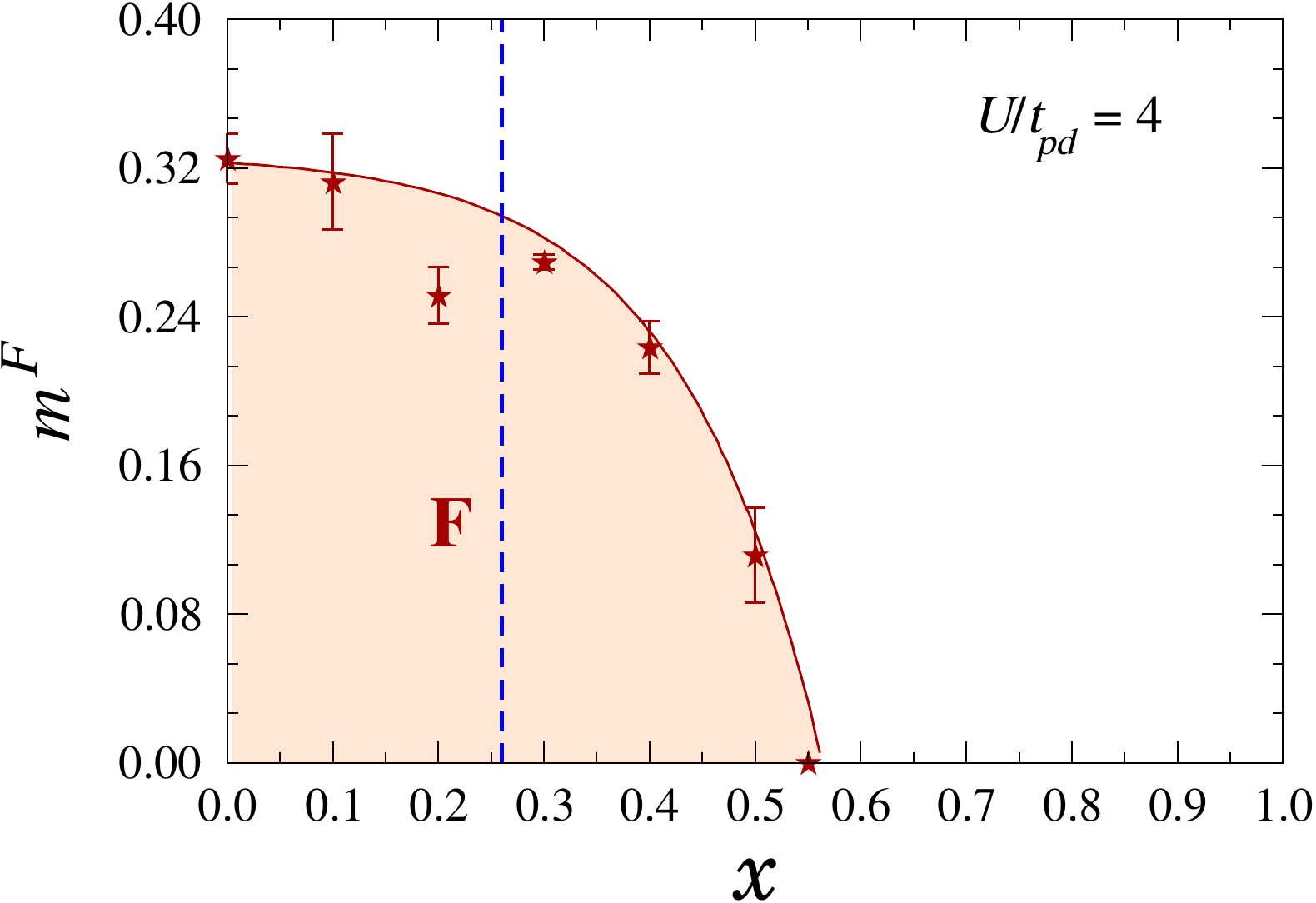}
\caption{(Color online) Global ground state ferromagnetic order
parameter $m^{\it F}$ as a function of dilution fraction, $x$. The error
bars are due to the uncertainties in the $1/L\to0$ extrapolation; see
Fig.\,\ref{fig:S00Huse}.
The solid curve is a guide to the eye for the magnetization, while the vertical dashed line marks the classical site-percolation threshold for the Lieb lattice.}
\label{fig:mFx} 
\end{figure}

Figure \ref{fig:S00beta} illustrates the behavior of the global uniform
structure factor $S(0,0)$ with the inverse temperature, 
$\beta=1/T$,
for three different concentrations of free sites.  In each case $S(0,0)$
approaches an asymptotic value for sufficiently large $\beta$,
reflecting the fact that, in an ordered phase, 
the correlation length is limited by the
finite size of the lattice.  These large $\beta$ values give
$S(0,0)$ at $T=0$ (for the given system size and
concentration) and are used for the scaling analysis of
Eq.\,\eqref{eq:Huseglobal}.  The outcome is depicted in
Fig.\,\ref{fig:S00Huse}.  For each concentration the
ground state magnetization in the thermodynamic limit
is obtained from the intercept with the
vertical axis ($1/L=0$).
These, in turn, are plotted as a function of the concentration in
Fig.\,\ref{fig:mFx}.

From Fig.\,\ref{fig:mFx} we see that the global magnetization decreases
steadily with increasing disorder and vanishes around $x_c\approx 0.55$,
a value more than twice as large as the classical site-percolation threshold for the
Lieb lattice,
$x_c^\text{(perc,Lieb)}\approx 0.26$ \cite{Tarasevich99,Oliveira19}.  
This clearly shows that the
disorder-induced transition is not purely geometric.

\begin{figure}[t!]
\centering
\includegraphics[scale=0.40]{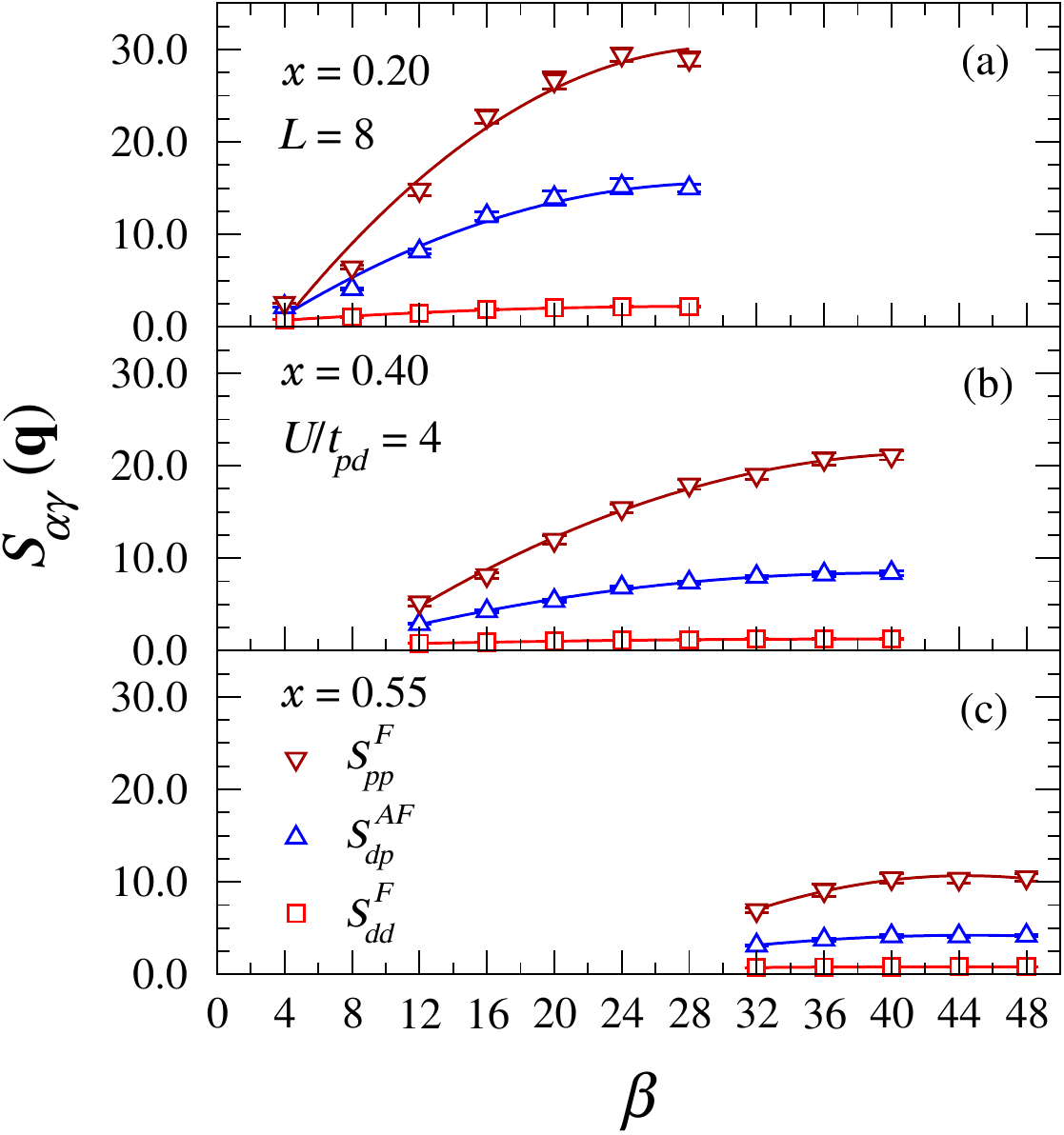}
\caption{(Color online) Orbitally-resolved contributions to the
structure factor from $d$-$d$ (squares), $d$-$p$ (triangles), and
$p$-$p$ (inverted triangles) correlations as functions of the inverse
temperature, for fixed $L=8$ and (a) $x=0.20$, (b)~$x=0.40$, and (c)
$x=0.55$. Solid lines are guides to the eye.}
\label{fig:Sagbeta}
\end{figure}

\begin{figure}[t!]
\centering
\includegraphics[scale=0.40]{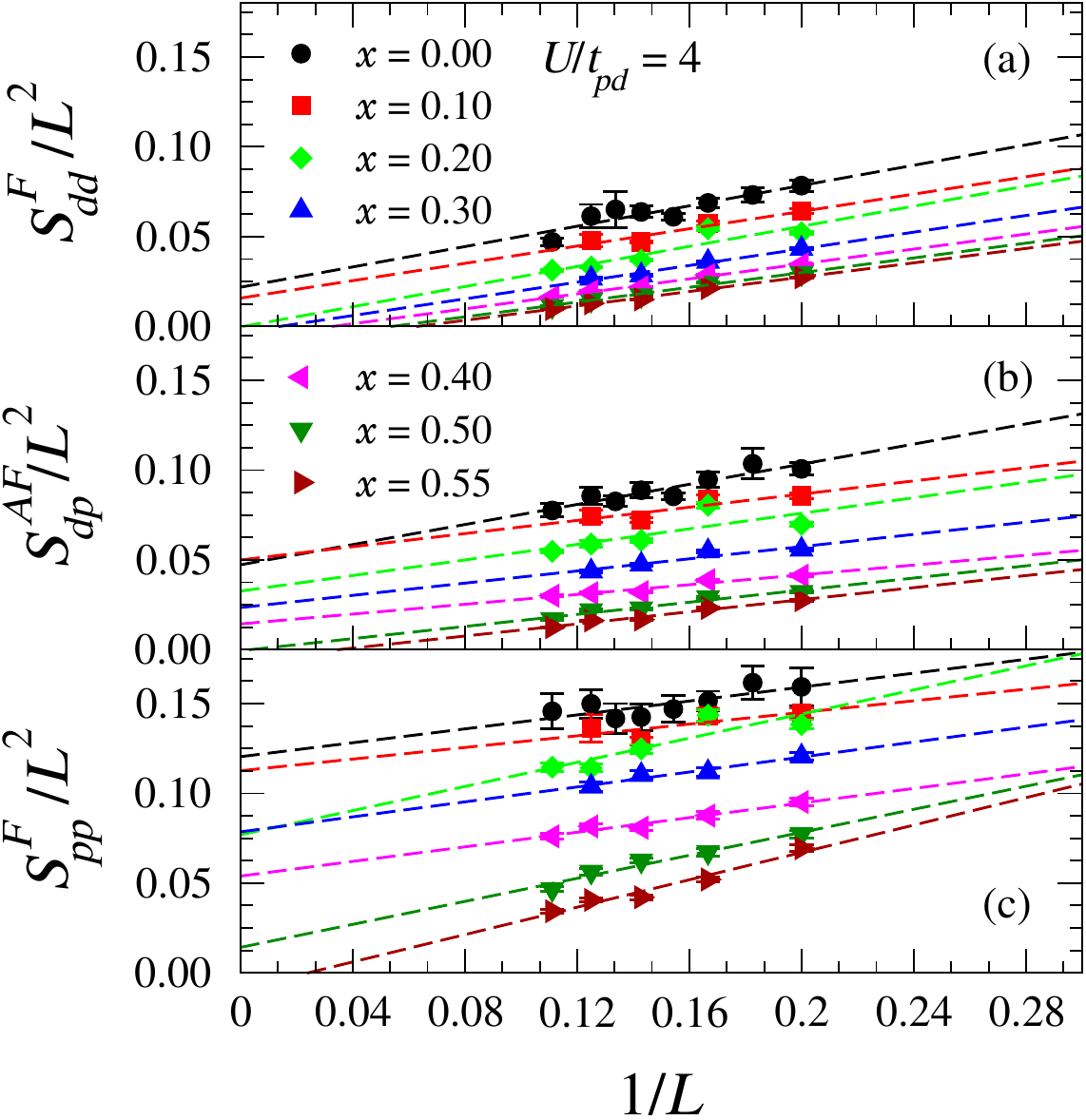}
\caption{(Color online) Finite-size scaling of the normalized,
orbitally-resolved structure factors of (a) $d$-$d$, (b) $d$-$p$, and
(c) $p$-$p$ contributions for different dilutions.}
\label{fig:SagHuse}
\end{figure}

\begin{figure}
\centering
\includegraphics[scale=0.28]{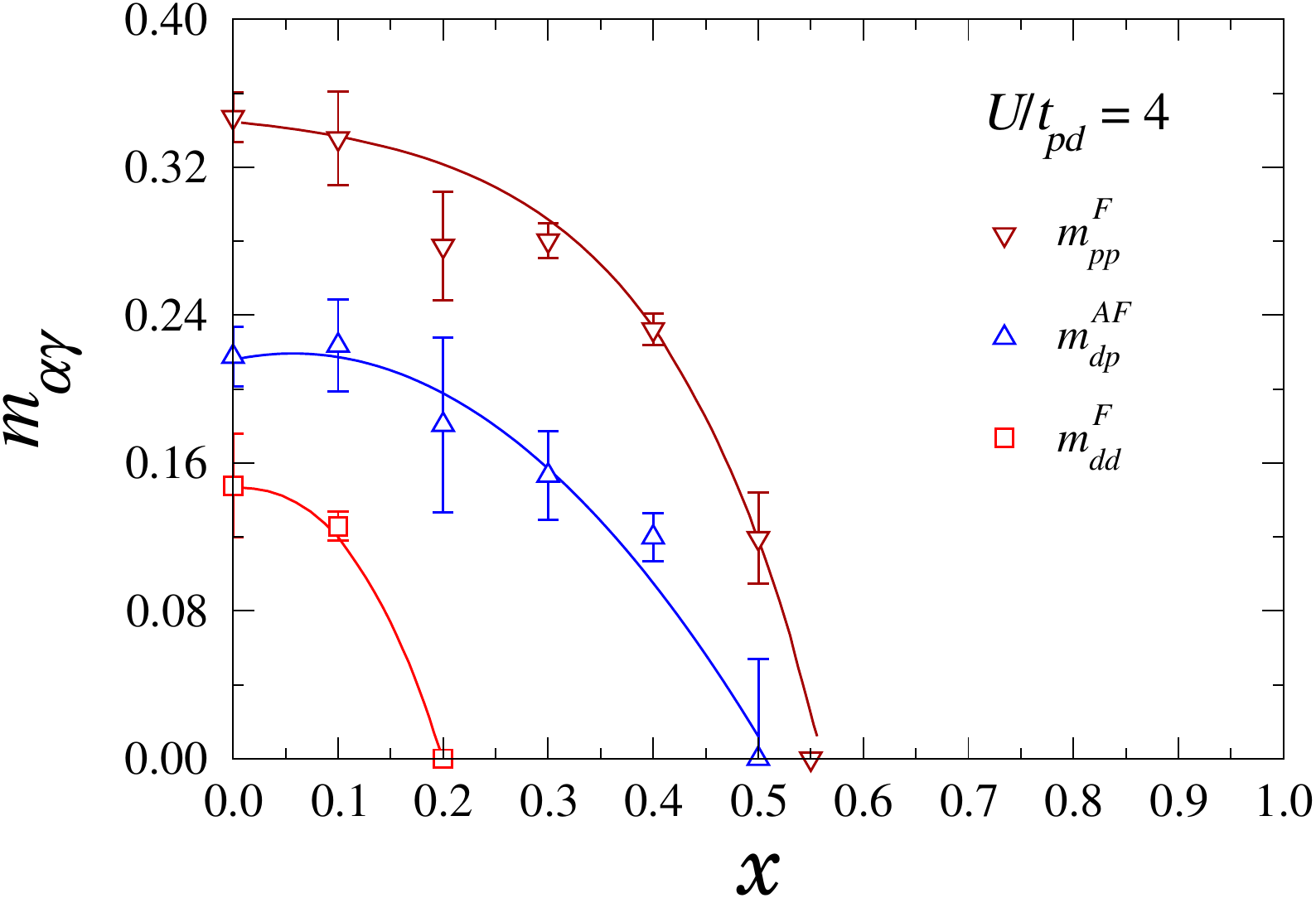}
\caption{(Color online) Extrapolated ($L\to\infty$) values of the
orbitally-resolved contributions to the order parameter, obtained from
the scaling of their structure factors; see Fig.\,\ref{fig:SagHuse}.
Solid lines are guides to the eye.}
\label{fig:magx} 
\end{figure}

In order to understand why the magnetic behavior on the diluted Lieb
lattice is more robust, in the sense that
$x_c > x_c^\text{(perc,Lieb)}$, 
we must examine the orbital-resolved order
parameters.  Figure \ref{fig:Sagbeta} shows 
their temperature dependence
for a given linear
system size.  We see that the dominant correlations between electrons on
$p$ orbitals are ferromagnetic, and so are those on $d$ sites; by
contrast, when one electron is on a $p$-site and the other on a
$d$-site, the correlations are antiferromagnetic, justifying
the form of Eq.~\ref{Eq:StrucFactor2}.
Following the
procedure adopted for the global structure factor, in
Fig.\,\ref{fig:SagHuse} we extrapolate the low temperature results
to $L\to\infty$.  The thermodymanic limit intercepts 
with the vertical axis are plotted in Fig.\,\ref{fig:magx}.

A strong coupling analysis for the clean system~\cite{Costa16}
attributes 
the  robustness of the $pp$
FM order parameter to the $p$ spins locking into triplets. 
In contrast, the
weakness of the $dd$ correlations originates in a shielding by these surrounding
triplets. 
This picture in fact persists to seemingly rather small 
values of $U/t_{pd}$, as a result
of the flatness of $p$-band, which makes the ratio
of the interaction to bandwidth large.  
The formation of such triplets seems to be only weakly affected if $U=0$ on all $d$-sites:
as discussed in Ref.\,\onlinecite{Costa16}, magnetic order persists even in this limiting case.
Upon random dilution, one should notice that the long-range behavior of $d$-sites ($m^{F}_{dd}$) is strongly suppressed for a small dilution strength, 
while magnetism is dominated by the $p$-sublattice, as shown
in Fig.\,\ref{fig:magx}:
the $pp$ contribution to the magnetization is much stronger than
those involving $d$-sites, both in intensity and in its resilience to
disorder, sustaining order well
beyond the classical percolation threshold.
It is worth noticing that the robustness of the long-range behavior of $p$-sites is due to their coupling to $d$-sites, which may restore the triplets even when $U=0$ on a given $p$-site.
As displayed in Fig.\,\ref{fig:magx}, long-range antiferromagnetic
correlations between $p$ and $d$-electrons (i.e., $m^{AF}_{dp}$) occur
even for $m^{F}_{dd}=0$, and has almost the same threshold as $m^{F}_{pp}$.
It therefore emphasizes the importance of $d$-electrons to global magnetism, and seems to be the key feature for the occurence of magnetism beyond the classical percolation limit.

\begin{figure}
\centering
\includegraphics[scale=0.32]{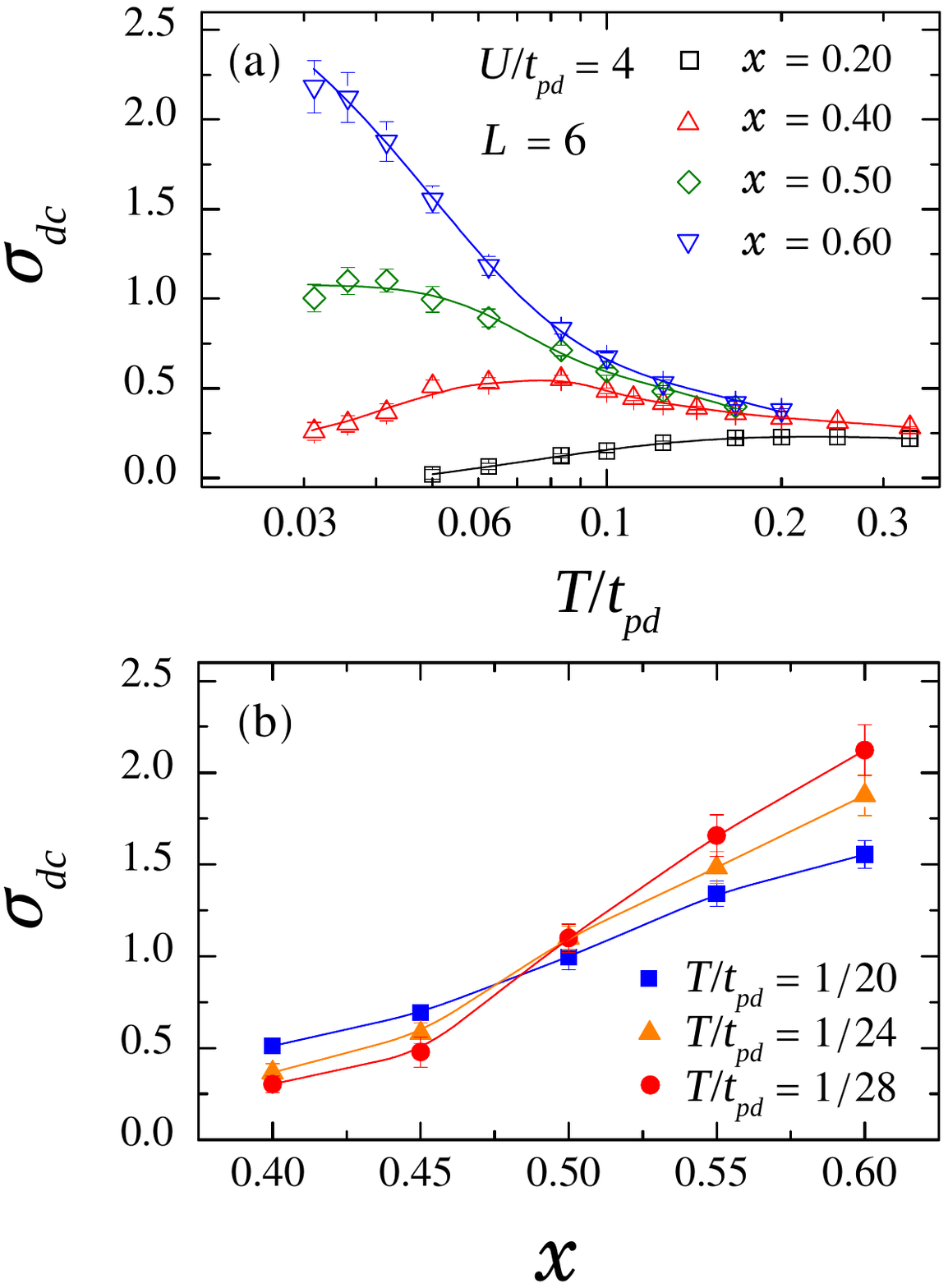}
\caption{(Color online) Conductivity (a) as a function of temperature, for
different dilutions, and (b) as a function of dilution, for different
temperatures. The solid lines are guides to the eye.  Both presentations
of the data suggest insulating behavior for $x<x_c\sim 0.5$ and
metallic behavior for $x>x_c\sim 0.5$.}
\label{fig:conductivity}
\end{figure}

Once the disorder threshold is exceeded, there are not enough 
strongly repulsive $U$-sites
to sustain an insulating state at half-filling, and we expect a 
metallic state to set in.
This can be checked with the aid of the conductivity, calculated
through Eq.\,\eqref{eq:sigma_dc}.  Figure \ref{fig:conductivity}\,(a)
shows the temperature dependence of $\sigma_{dc}$ for different
disorder concentrations, while Figure \ref{fig:conductivity}\,(b) shows
$\sigma_{dc}$ as a function of concentration, for different
temperatures.  Two distinct regimes are clearly identified:
insulating, when $\sigma_\text{dc}$ decreases as the temperature
decreases, and metallic, when $\sigma_\text{dc}$ increases as the
temperature decreases.
One can roughly estimate that the change in behavior
occurs at $x_c^{(\sigma_\text{dc})}=0.50\pm0.03$, which is consistent
with the results suggesting a change in magnetic behavior at
the same $x_c$.
The transition across
$x_c$ is therefore from an insulating ferrimagnetic phase
to a metallic paramagnetic one.


\begin{figure}
\centering
\includegraphics[scale=0.28]{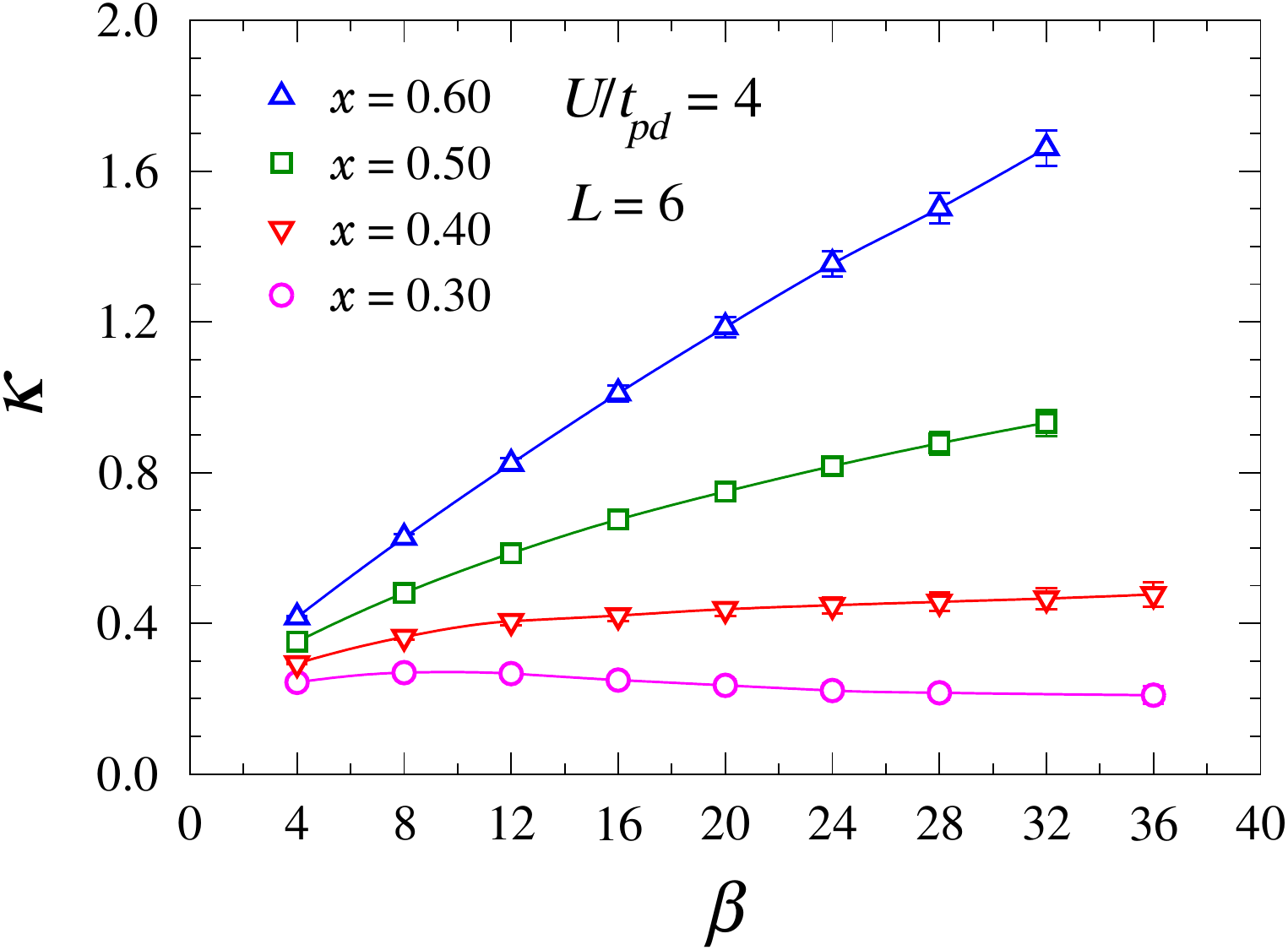}
\caption{(Color online) Global compressibility as a function of the
inverse temperature, for different dilution concentrations, $x$. The
solid lines are guides to the eye.}
\label{fig:kappa-global} 
\end{figure}

\begin{figure}
\centering
\includegraphics[scale=0.35]{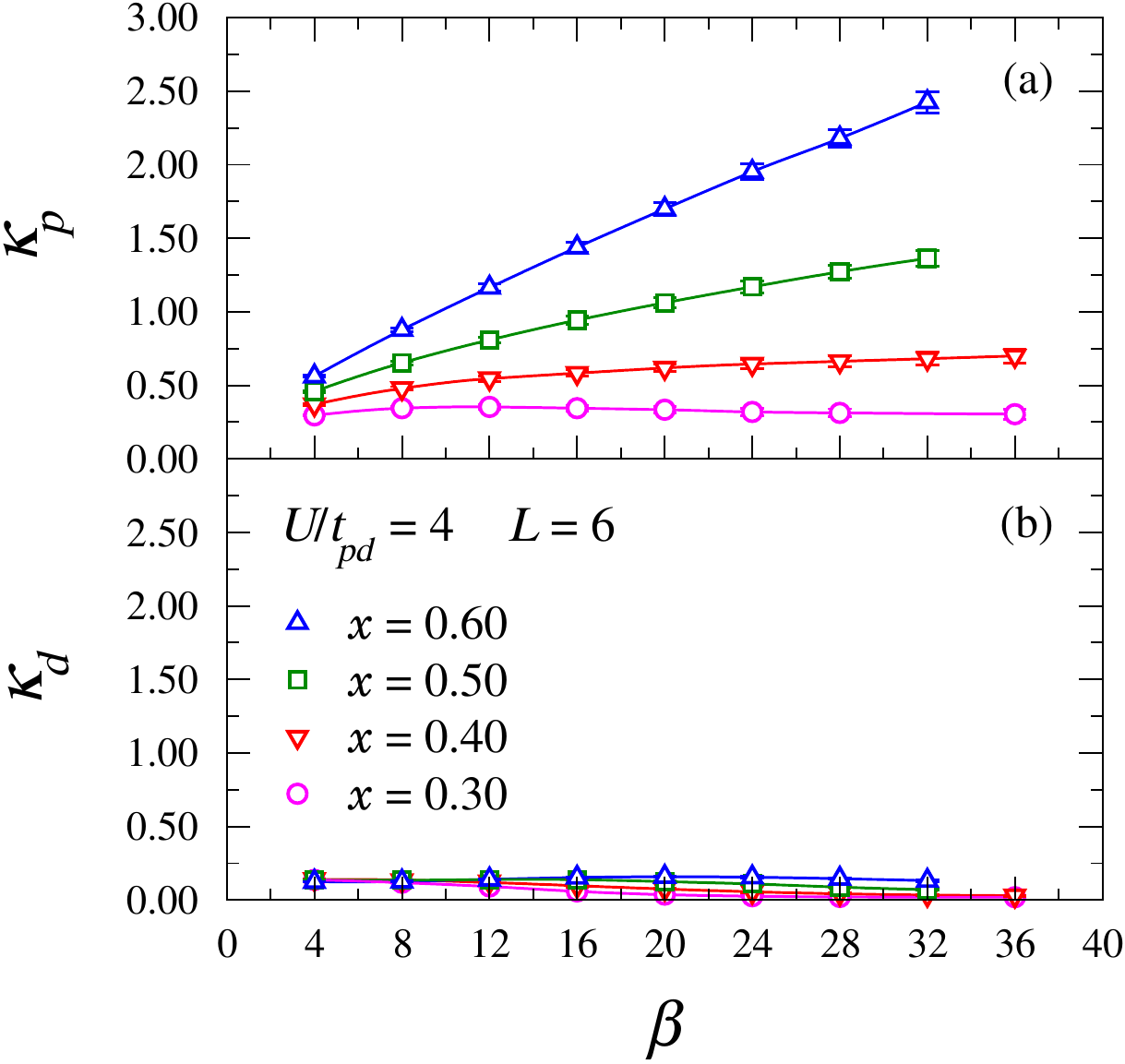}
\caption{(Color online) Site-resolved compressibility as a function of
the inverse temperature, for different dilution concentrations, $x$: (a)
compressibility on $p$ sites, and (b) compressibility on $d$ sites.
Solid lines are guides to the eye.}
\label{fig:kappa-ind} 
\end{figure}

Further evidence in favor of the insulator-metal transition is provided
by the compressibility, Eq.\,\eqref{eq:kappa}.  Figure
\ref{fig:kappa-global} shows the \emph{global} compressibility, 
the change in the \emph{overall} density with chemical
potential.  We see that for $x=0.3$ and 0.4 the system has a small,
temperature independent value of $\kappa$,
while for $x=0.5$ and 0.6, $\kappa$
increases with $\beta$.  In Figure \ref{fig:kappa-ind} the
compressibility is broken into individual contributions from $p$ and $d$
sites, and we see that the dominant behavior comes from $p$ sites, which
are the ones forming the flat band.
 That is, $x_c$ defines a value above which the $p$ sites become weakly compressible. It is also worth noticing that for $x>x_c$, $\kappa_p$ seems to diverge at low temperatures, a behavior already present  in the noninteracting case due to the dispersionless middle band [see, e.g., Fig.\,\ref{fig:Lieb}\,(c)]. Thus, for $x>x_c$, the transport properties resemble those for the noninteracting one, despite the presence of electron-electron interactions on a subset of sites.

The following picture emerges from the combination of data for the
magnetic structure factor, conductivity, and compressibility:
The low temperature insulating ferrimagnetic state can accommodate 
extra electrons on the
$U=0$ sites at low energetic cost, provided there are not too many of
them.  As dilution increases, more $U=0$ $p$ sites become
available, and the system becomes fully compressible.
One should also notice that, since there is an
energetic cost to break the triplets formed by spins on $p$ sites, the compressibility is reduced as the temperature decreases, for $x \lesssim 0.30$, and 
grows faster for $x \gtrsim 0.40$.

\section{Conclusions}
\label{sec:conc}

Studies of the Periodic Anderson Model~\cite{Assaad02b,Costa18,Zhang19}
and of the single band Hubbard model with random dilution
of the on-site interaction~\citep{Ulmke98b,Litak00,Hurt05,Pradhan18,Mondaini08}
have provided opportunities for the exploration of magnetic order 
in systems possessing both sites where moments form and those
for which charge fluctuations are allowed.
By considering dilution on the repulsive Hubbard model on a Lieb
lattice, in which the on-site repulsion $U$ is switched off on a
fraction $x$ of sites, we have established some 
new features of this problem.

In particular, although studies of the diluted square lattice Hubbard
model suggested the critical concentration for $U=0$ sites is
{\it less than} the percolation value, we have shown
here that on the Lieb lattice 
magnetic order is {\it more robust} than
one might expect from percolation arguments:
the percolation threshold, $x_c$, is higher than the one solely
determined by the geometry of the lattice, 
$x_c^\text{(perc,Lieb)}$.
While a dynamic (i.e.\,~interaction-driven) influence on $x_c$ had already
been noted \cite{Hurt05,Mondaini08} for the attractive Hubbard model at
half-filling (which has a corresponding behavior in the repulsive case),
there the classical percolation threshold provides an \emph{upper} bound
to the quantum case. 
A second observation is that,
simultaneously with the `percolative' magnetic transition, the system
undergoes an insulator to metal transition, as evidenced by both the
dc-conductivity and the compressibility.  These properties are a direct
consequence of the flat $p$-band displayed by the non-interacting Lieb
lattice.  Our results therefore show that disordered quantum itinerant
systems display a non-trivial interplay between dynamics and lattice
geometry leading to features with no counterpart in classical systems.

Previous works~\cite{Ulmke98b,Litak00,Hurt05,Pradhan18,Mondaini08}
have suggested that
the presence of two regimes, one at strong coupling 
where 
$x_c\sim x_c^\text{(perc,sq)}$,
and one at weaker coupling where
$x_c < x_c^\text{(perc,sq)}$,
might be connected to the 
two distinct physical pictures for the origin of
antiferromagnetism (AF) in the half-filled single-band square lattice
Hubbard model.  
For large $U/t$ one thinks of a Mott insulating state in which
a superexchange interaction $J = 4t^2/U$ couples neighboring spins. 
A (quantum) Heisenberg spin de\-scrip\-tion is appropriate in this regime.
On the other hand, at weak coupling, AF can be viewed as arising from
a spin-density wave instability driven by Fermi surface nesting.
In this case, the $U=0$ band structure and
electron itinerancy play a central role.
Our work suggests that, although a criterion for $x_c$ 
based purely on the strength of $U/t$ might be correct for
a single band model, 
a more complex picture is necessary to understand the 
multiple band case.
Specifically, the orbitally-resolved magnetic order parameters
$m_{\alpha\gamma}$
must be analyzed, and might vanish at very different dilution
fractions.
Finally, it would also be interesting to verify whether the metal-insulator and the ferrimagnetic transitions always take place concomitantly, or may occur at different regions of the parameter space, e.g.\, if turns out that $x_{c}^\mathrm{Fer}\neq x_{c}^\mathrm{Ins}$, one could have a ferrimagnetic metal or a nonmagnetic Mott insulating state. Further work is needed to clarify this interesting issue.

\section*{ACKNOWLEDGMENTS}
The work of RTS was supported by the grant DE‐SC0014671 funded by
the U.S. Department of Energy, Office of Science.
Financial support from the Brazilian Agencies
CAPES, CNPq,  FAPERJ and FAPEPI is also gratefully acknowledged.
The authors are grateful to T. Paiva and S.L.A. de Queiroz for discussions.
N.C.C. acknowledges PRACE for awarding him access to Marconi at CINECA, Italy.


\bibliography{Lima_Lieb2}
\end{document}